# Bilayer Membrane in Confined Geometry: Interlayer Slide and Steric Repulsion.

S.V. Baoukina[1], S.I. Mukhin[1]

[1] Theoretical Physics Department, Moscow Institute for Steel and Alloys, Leninsky Pr., 4, 119049 Moscow, Russia

## Abstract

We derived free energy functional of a bilayer lipid membrane from the first principles of elasticity theory. The model explicitly includes position-dependent mutual slide of monolayers and bending deformation. Our free energy functional of liquid-crystalline membrane allows for incompressibility of the membrane and vanishing of the in-plane shear modulus and obeys reflectional and rotational symmetries of the flat bilayer. Interlayer slide at the mid-plane of the membrane results in local difference of surface densities of the monolayers. The slide amplitude directly enters free energy via the strain tensor. For small bending deformations the ratio between bending modulus and area compression coefficient, $K_b/K_A$, is proportional to the square of monolayer thickness, $h$. Using the functional we performed self-consistent calculation of steric potential acting on bilayer between parallel confining walls separated by distance $2d$. We found that temperature-dependent (T) curvature $\alpha$: $\alpha \propto T^2/K_b d^4$, at the minimum of confining potential is enhanced four times for a bilayer with slide as compared with a unit bilayer. We also calculate viscous modes of bilayer membrane between confining walls. Pure bending of the membrane is investigated, which is decoupled from area dilation at small amplitudes. Three sources of viscous dissipation are considered: water and membrane viscosities and interlayer drag. Dispersion relation gives two branches $\omega_{1,2}(q)$. Confinement between the walls modifies the bending mode $\omega_1(q)$ with respect to membrane in bulk solution. Four dependencies are obtained: $\omega_1 \sim -i\alpha d^3 q^2/\eta_w$; $-iK_b d^3 q^6/\eta_w$; $-iK_A q^2/b_s$ and $-iK_A/\eta_m h$ for the consecutive intervals of wave vector q: $q<<(\alpha/K_b)^{1/4}$; $(\alpha/K_b)^{1/4}<<q<<1/d$ ; $1/d<<q<<(b_s/h\eta_m)^{1/2}$ and $(b_s/h\eta_m)^{1/2}<<q$. Simultaneously, interlayer slipping mode $\omega_2$, damped by viscous drag, remains unchanged by confinement: $\omega_2 \sim -iK_A q^2/b_s$ and $\omega_2 \sim -iK_b q^3/\eta_w$ for $q<<1/d$ and $1/d<<q$, respectively.

## 1 Introduction

Cell membrane is characterized by complex structural and dynamical properties [1,2,3]. Theoretical modeling and description of lipid membranes is of great fundamental and practical interest and has long enough history. Phenomenological model introduced by Helfrich [4] treated lipid membrane as a single sheet with bending rigidity and spontaneous curvature. This model was later used for calculation of frequency spectrum of membrane in water solution [5] and for investigation of steric interactions of membranes in multilayer systems [6]. Bilayer structure of lipid membrane was analyzed by Evans and Yeung [3, 7], who considered dynamic coupling between the monolayers and interlayer slide. Allowing for the coupling between local curvature and local densities of lipids within the monolayers the frequency spectrum of membrane in the bulk water was recalculated [8]. Afterwards, viscous modes of a bilayer adhering to a substrate were found [9] using density-difference model [8], supplemented with binding potential [10].

In this paper we derive new free energy functional of a bilayer membrane with interlayer slide. Interlayer slide function, membrane stretching and bending amplitude enter directly the strain tensor of the membrane. Our functional is distinct from but can be reduced (as particular case) to density-difference model used in [8, 9]. We study dynamics of bilayer membrane in water solution confined between parallel walls as a step towards understanding inter-membrane interactions. The effect of confinement is modeled by steric potential [11].

In Section 2 we introduce an anisotropic elastic moduli tensor containing initially 21 independent components. The reflection and rotation symmetries of the flat bilayer reduce the number of components to 5. Next, we impose zero shear stress modulus and incompressibility constraint. We restrict ourselves only to the case of small bending deformations and exclude the corresponding strain and elastic tensor components. Thus, the number of independent components of elastic tensor in the free energy functional is reduced to two. The derived free energy functional of a bilayer membrane contains three fields describing area dilation and bending deformation coupled to interlayer slide.

In Section 3 a parabolic steric potential acting on the membrane between confining walls is introduced. We calculate self-consistently the curvature of the confining potential at its minimum. We evaluate the curvatures of the steric potential for a bilayer with slide and for a unit bilayer.

In Section 4 we use the derived functional to study dynamical properties and dissipative mechanisms of the bilayer membrane in water solution confined between parallel walls. We investigate only pure bending deformations of the membrane (zero total lateral stretching), which decouple from area dilation. Velocity field in the surrounding water is found by solving Stockes equations for incompressible fluid. Fluid velocity vanishes at the walls. Equations of motion are determined as boundary conditions on the membrane surfaces by requirement of force balance neglecting inertial effects. Three sources of dissipation are included into dynamic equations: water and membrane viscosities and interlayer drag.

In the last Section of the paper we discuss limitations and possible improvements of our model and correspondence with earlier results [9]. In Appendix A static behavior of membrane in axial-symmetric case is studied. Analytical solutions are obtained for a circular membrane bent by external pressure. Membrane bending, interlayer slide and lateral stress distribution are found as functions of pressure across the membrane. In Appendix B we rederive the dispersion relation [8] for a membrane in the bulk water solution using our free energy functional.

## 2 Free energy functional

Free energy density of anisotropic medium can be written to the lowest order in elastic strain tensor as [12,13]:

$$F = \frac{1}{2} \cdot \lambda_{iklm} \cdot u_{ik} \cdot u_{lm},  \qquad (1)$$

where a summation over the repeated indices *i, k, l, m* is performed. Indices *i, k, l, m* acquire values 1, 2, 3, numerating the space axis *x, y, z* appropriately. Here $u_{ik}$ is the strain tensor, $\lambda_{iklm}$ is the elastic (modulus) tensor. By definition the elastic tensor is symmetric under the exchange $i \leftrightarrow k$, $l \leftrightarrow m$ and $i,k \leftrightarrow l,m$:

$$\lambda_{iklm} = \lambda_{kilm} = \lambda_{ikml} = \lambda_{lmik},$$

and has 21 independent coefficients.

Allowing for (1), the (symmetric) stress tensor $\sigma_{ik}$ is defined as:

$$\sigma_{ik} = \frac{\partial F}{\partial u_{ik}} = \lambda_{iklm} \cdot u_{lm}. \qquad (2)$$

In a symmetric medium there is a correlation between different components $\lambda_{iklm}$ and the number of independent elements of the tensor of elastic modulus is reduced.

Let us introduce Cartesian coordinate system with the z-axis perpendicular to the unperturbed (flat) membrane plane, and with the monolayer interface (i.e. bilayer midplane) positioned in the (x,y)-plane (at z=0). The membrane thickness is equal to *2h*, and the flat membrane is modeled as a thin bilayer plate bound by the *z=-h* and *z=h* planes with in-plane linear dimension *R>>2h*. The *(x,y)* plane is a plane of reflection symmetry. This implies that under a transformation x→x, y→y, z→-z the free energy must be invariant. Therefore all the components $\lambda_{iklm}$ with odd number of *z*-indices are equal to zero [9]. Membrane can be considered laterally isotropic. Then z-axis is an axis of rotational symmetry. Thus the expression for elastic energy density, F, reduces to the following [10]:

$$F = \frac{1}{2} \cdot \lambda_{xxxx} \cdot (u_{xx}^2 + u_{yy}^2) + \frac{1}{2} \cdot \lambda_{zzzz} \cdot u_{zz}^2 + \lambda_{xxyy} \cdot u_{xx} \cdot u_{yy} + (\lambda_{xxxx} - \lambda_{xxyy}) \cdot u_{xy}^2$$
$$+ \lambda_{xxzz} \cdot (u_{xx} \cdot u_{zz} + u_{yy} \cdot u_{zz}) + 2 \cdot \lambda_{xzxz} \cdot (u_{xz}^2 + u_{yz}^2) \qquad (3)$$

Assuming that the membrane is in liquid state, we require that the in-plane shear modulus (the coefficient in front of $u_{xy}^2$) vanishes, and thus obtain: $\lambda_{xxxx} = \lambda_{xxyy}$. Hence, expression (3) farther simplifies and acquires the form:

$$F = \frac{1}{2} \cdot \lambda_{xxxx} \cdot (u_{xx} + u_{yy})^2 + \frac{1}{2} \cdot \lambda_{zzzz} \cdot u_{zz}^2 + \lambda_{xxzz} \cdot (u_{xx} \cdot u_{zz} + u_{yy} \cdot u_{zz}) +$$
$$+ 2 \cdot \lambda_{xzxz} \cdot (u_{xz}^2 + u_{yz}^2) \qquad (4)$$

Let an external force applied perpendicular to the membrane plane induce a small bending deformation along the *z*-axis. Allowing for a typical experimental situation, we consider a thin membrane with the ratio of its thickness to lateral linear dimension (effective radius) of the order of $10^{-3}$. Hence we neglect the applied external stresses on the top and bottom membrane surfaces with respect to the internal lateral stresses in it. Due to the smallness of the membrane thickness, zero stresses on the surface also vanish in the bulk of the membrane. So we impose the following condition usually implied for the thin plates [12]:

$$\sigma_{xz}(\mathbf{r}) = \sigma_{yz}(\mathbf{r}) = \sigma_{zz}(\mathbf{r}) \equiv 0 ; \tag{5}$$

where **r** spans over the membrane's bulk. In accord with (2), (4), these components of the stress tensor are related to the strain tensor components as follows:

$$\sigma_{xz} = 4 \cdot \lambda_{xzxz} \cdot u_{xz}, \quad \sigma_{yz} = 4 \cdot \lambda_{yzyz} \cdot u_{yz}, \tag{6}$$

$$\sigma_{zz} = \lambda_{xxzz} \cdot (u_{xx} + u_{yy}) + \lambda_{zzzz} \cdot u_{zz}. \tag{7}$$

Combining the relations (5) and (7), we find:

$$u_{zz} = -\frac{\lambda_{xxzz}}{\lambda_{zzzz}} \cdot (u_{xx} + u_{yy}). \tag{8}$$

It is interesting to mention that as follows from (6) fulfillment of the first two conditions (5) requires vanishing of the strain tensor components $u_{xz}$ and $u_{yz}$. Vanishing of these components would also correspond to infinite elastic moduli $\lambda_{xzxz}$ and $\lambda_{yzyz}$.

Condition (5) permits to omit the terms containing $u_{xz}$ and $u_{yz}$ in (4). Also using (8) and expressing $u_{zz}$ via $u_{xx} + u_{yy}$ in (4), we find the expression for the free energy density:

$$F = \frac{1}{2} \cdot \left( \lambda_{xxxx} - \frac{\lambda_{xxzz}^2}{\lambda_{zzzz}} \right) \cdot (u_{xx} + u_{yy})^2 . \tag{9}$$

In addition, we impose the "incompressibility" condition, i.e. the constancy of the bulk density of the membrane:

$$u_{xx} + u_{yy} + u_{zz} = 0 . \tag{10}$$

The condition (10) is satisfied simultaneously with (8) if: $\lambda_{zzzz} = \lambda_{xxzz}$.

Finally, the free energy density is written as:

$$F = \frac{1}{2} \cdot K_1 \cdot (u_{xx} + u_{yy})^2 , \tag{11}$$

where $K_1$ denotes a superposition of anisotropic elastic moduli: $K_1 = (\lambda_{xxxx} - \lambda_{zzzz})$.

In the linear approximation for the strain tensor one has:

$$u_{ik} = \frac{1}{2} \cdot \left( \frac{\partial u_i}{\partial x_k} + \frac{\partial u_k}{\partial x_i} \right), \tag{12}$$

where $u_i$ is the *i*-th component of the distortion field.

In order to introduce the essentials of our model in a simplest way, we limit the following discussion to the small bending amplitude case, i.e. we impose condition: $|u_z| \ll h$, where $u_z(\mathbf{r})$ is the z-component of displacement, describing the deformed membrane. Also, we shall neglect the z-dependence of the component $u_z(\mathbf{r})$ in the thin plate approximation [12], thus defining the "shape"-function $\xi(x,y) \approx u_z(\mathbf{r})$ independent of the depth z. Substituting $\xi(x,y)$ in the definition (12) and thereafter into the relations (6) and conditions (5), one obtains the following partial differential equations:

$$\frac{\partial u_x}{\partial z} = -\frac{\partial \xi}{\partial x},$$

$$\frac{\partial u_y}{\partial z} = -\frac{\partial \xi}{\partial y}. \tag{13}$$

Now, while integrating equations (13), let us introduce two functions: (inhomogeneous) lateral stretching of the membrane **a**(x,y) and in-plane slide $\pm$**f**(x,y) of the lower (z<0) and upper (z>0) monolayers at the mid-plane z=0 of the membrane. Thus, the in-plane distortions $u_x$ and $u_y$ of a bilayer membrane have the following form:

$$u_x = -z \cdot \frac{\partial \xi(x,y)}{\partial x} + (\Theta(z) - \Theta(-z)) \cdot f_x(x,y) + a_x(x,y),$$

$$u_y = -z \cdot \frac{\partial \xi(x,y)}{\partial y} + (\Theta(z) - \Theta(-z)) \cdot f_y(x,y) + a_y(x,y), \tag{14}$$

where the step-function is defined as: $\Theta(z>0) \equiv 1$ and $\Theta(z<0) \equiv 0$, and the choice of the sign of $\Theta$ and of its argument is made for the farther convenience.

It is worth emphasizing here the limitations of validity of the relations (14). The expressions (14) are clearly distinct from the usual expressions for thin plates [12]. In the latter case the displacements $u_x$ and $u_y$ are set to zero at z=0, implying the presence of neutral (not stretched) surface at the mid-plane of the plate in the small bending approximation: $\xi \ll h$ [12]. It will be shown in Appendix A (see equation A.11) that the second term in (14) is of the same order as the first one: $f_{x,y} \sim h \cdot \xi / R$, where R is effective radius of the membrane. Small bending approximation is justified when the quadratic in $\xi$ term is negligibly small with respect to linear terms in the expressions for in-plane distortions $u_x$ and $u_y$ : $O(\xi^2/R) \ll h \cdot \xi / R$. The latter condition is fulfilled as long as $\xi \ll h$. On the other hand, for strongly bent thin plate the $\xi^2$–term dominates over $\xi$–term and thus higher order terms should be added on the right hand side of equations (14).

Let us now discuss the physical meaning of the expressions (14). The membrane stretching **a**(x,y) defines position-dependent shift of the neutral surface (along z coordinate) while slide function **f**(x,y) multiplied by the step-functions lead to the splitting of this neutral surface into two surfaces belonging to upper and lower monolayers. These surfaces are determined from the conditions: $u_x(x,y,z) \equiv 0$ and $u_y(x,y,z) \equiv 0$. The function **f**(x,y) provides additional degree of freedom in comparison with a bilayer without slide (or a single monolayer). Under the condition of zero total lateral stretching (i.e. pure bending deformation, $\mathbf{a} \equiv 0$) the presence of the function $\mathbf{f}$ means that the neutral surface splits into two such surfaces located in each monolayer symmetrically with respect to the mid-plane z=0. The total amplitude of mutual interlayer slide at each point *x,y* of the mid-plane is then given by $2\mathbf{f}(x,y)$, which signifies discontinuity of in-plane distortions $u_x$ and $u_y$ across the mid-plane z=0. In the opposite case: $\mathbf{f} \equiv 0$ the monolayers are coupled together (no interlayer slide) and the distortion field is the sum of bending and stretching (for small deformations), the latter being continuous across the mid-pane z=0. In general, the distortion field (14) includes bending, stretching and mutual interlayer slide.

Substituting (14) into (12), we proceed to determine the strain tensor components:

$$u_{xx} = -z \cdot \frac{\partial^2 \xi(x,y)}{\partial x^2} + (\Theta(z) - \Theta(-z)) \cdot \frac{\partial f_x(x,y)}{\partial x} + \frac{\partial a_x(x,y)}{\partial x},$$

$$u_{yy} = -z \cdot \frac{\partial^2 \xi(x,y)}{\partial y^2} + (\Theta(z) - \Theta(-z)) \cdot \frac{\partial f_y(x,y)}{\partial y} + \frac{\partial a_y(x,y)}{\partial y},$$ (15)

and $u_{zz}$ can be expressed via $u_{xx}$ and $u_{yy}$ using (8).

It is important to mention that (in accord with definition (12)) the addition of discontinuous terms $f_{x,y} \cdot \Theta(\pm z)$ in (14) leads to non-vanishing contributions to the $u_{xz}$ and $u_{yz}$ components of the strain tensor proportional the Dirac's delta function $\delta(z)$. Allowing for the energy cost-free (static) inter-monolayer slide, we discard these contributions to $u_{xz}$ and $u_{yz}$, thus keeping the latter equal to zero in accord with conditions (5), (6). The free energy functional of the whole membrane $F_v$ is obtained by an integration of the free energy density $F$ over the membrane's volume stepwise: first over the thickness coordinate (-h<z<h), and then over the membrane plane $\{x,y\}$. Using expressions (11) and (15), we finally find:

$$F_v = \frac{K_1}{2} \cdot \int (u_{xx} + u_{yy})^2 dV = \frac{K_1}{2} \cdot \left\{ \frac{2h^3}{3} \iint (\tilde{\nabla}^2 \xi)^2 dxdy \right.$$
$$\left. - 2h^2 \cdot \iint (\tilde{\nabla}^2 \xi) \cdot (\tilde{\nabla} \cdot \mathbf{f}) dxdy + 2h \cdot \iint ((\tilde{\nabla} \cdot \mathbf{a})^2 + (\tilde{\nabla} \cdot \mathbf{f})^2) dxdy \right\}$$ (16)

Here the tilde refers to two-dimensional differentiation ($\tilde{\nabla} = \left\{ \frac{\partial}{\partial x}, \frac{\partial}{\partial y} \right\}$).

Actually, equation (16) is quite remarkable. The mean curvature of the interlayer surface *H*, is expressed as follows:

$$\tilde{\nabla}^2 \xi = \frac{\partial^2 \xi}{\partial x^2} + \frac{\partial^2 \xi}{\partial y^2} \cong 2 \cdot H .$$ (17)

Therefore the first term on the right hand side gives effective bending energy, i.e. extrinsic curvature-bending energy functional, $F_c$, of the "standard" form [4, 12]:

$$F_c = \frac{K_b}{2} \cdot \int (2 \cdot H - c_0)^2 dS, \qquad (18)$$

with zero spontaneous curvature $c_0$. Here $K_b$ is bending rigidity (modulus). Comparing (16) and (18), we find: $2h^3 K_1 / 3 = K_b$.

The last term in (16) accounts for elastic energy of area dilation with area compression coefficient defined as $K_A = 2h \cdot K_1$. In general, local relative area dilation $\Delta S / S$ equals $u_{xx} + u_{yy}$ [12]. According to equation (15), the relative area dilation is given by $\tilde{\nabla} \cdot \mathbf{a}$, while the difference of relative area dilations between the monolayers is given by $2 \cdot (\tilde{\nabla} \cdot \mathbf{f})$. Hence, the $(\tilde{\nabla} \cdot \mathbf{a})^2$ term in (16) arises due to continuous (across the monolayers interface $z=0$) lateral stretching of the membrane, which leads to the change in average lipid density. The $(\tilde{\nabla} \cdot \mathbf{f})^2$ term represents the energy of local area difference of the monolayers (area-difference elasticity [2]), which is equivalent to difference of lipid densities in monolayers (density-difference model [8]). In principle, this energy is not related to the presence of neutral surfaces within the monolayers (at large membrane stretching/compression there are no neutral surfaces, which would obey $u_x \equiv u_y \equiv 0$, see expression(14)). As apparent from equation (16), the relation between bending and area compression coefficients (see [2]) $K_b / K_A \sim h^2$ comes out naturally in our derivation.

Next, the second term on the right hand side of equation (16) expresses coupling between bending deformation and interlayer slide producing local area dilation difference between monolayers. Note, that bending is decoupled from (continuous) area dilation, caused by lateral stretching, in the lowest order approximation. Due to hydrophobic effect the monolayers, while sliding, are forced to stick together and to follow the same shape defined by $\xi(x,y)$ on the monolayer interface. Mutual interlayer slide along the interface leads to relaxation of stretching/compression of the monolayers caused by bending deformation, and thus permits the free energy decrease.

Finally, our free energy functional is invariant to transversal slide of monolayers, such that div($\mathbf{f}$)=0. Hence, the energy does not change under a mutual rotation of the monolayers (as a whole) or a position independent shift of one of the monolayers with respect to the other.

We will consider pure bending deformations of the membrane with no overall stretching Therefore, we require the lateral strain integrated over the thickness to be zero in each point of the membrane. This imposes restriction on the form of $u_x$ and $u_y$: the function $\mathbf{a}(x,y)$ should be equal to zero in every point of the bilayer. Hence, this function is omitted everywhere below. Then strain tensor components can be written as:

$$u_{xx} = -z \cdot \frac{\partial^2 \xi}{\partial x^2} + (\Theta(z) - \Theta(-z)) \cdot \frac{\partial f_x}{\partial x},$$

$$u_{yy} = -z \cdot \frac{\partial^2 \xi}{\partial y^2} + (\Theta(z) - \Theta(-z)) \cdot \frac{\partial f_y}{\partial y}, \qquad (19)$$

and $u_{zz}$ again can be expressed via $u_{xx}$ and $u_{yy}$ using (8).

The free energy functional of the membrane acquires the form:

$$F_v = \frac{K_1}{2} \cdot \int (u_{xx} + u_{yy})^2 dV = \frac{K_1}{2} \cdot \left\{ \frac{2h^3}{3} \iint (\tilde{\nabla}^2 \xi)^2 dxdy \right.$$
$$\left. - 2h^2 \cdot \iint (\tilde{\nabla}^2 \xi) \cdot (\tilde{\nabla} \cdot \vec{f}) dxdy + 2h \cdot \iint (\tilde{\nabla} \cdot \vec{f})^2 dxdy \right\} \quad (20)$$

To study the properties of the functional (20) in detail a simple problem with cylindrically symmetric deformation is discussed in Appendix A. Equilibrium state of the membrane is defined by the Euler-Lagrange equations, which are obtained by equating to zero the first variational derivatives of the elastic energy functional $F(\xi, f)$ with respect to the functions $\xi(r)$ and $f(r)$.

## 3 Confining potential for a bilayer with slide

Direct influence of confined geometry on the membrane behavior manifests itself in the reduction of the manifold of accessible membrane conformations. Steric interactions of membrane with confining walls (see Figure 1) can be modeled [11] by introduction of an extra potential energy $W$ dependent on the bending amplitude: $W = \frac{\alpha}{2} \xi^2$. The free energy functional (20) appended with confining potential, $W$, acquires the form:

$$F_v = \frac{K_1}{2} \cdot \int (u_{xx} + u_{yy})^2 dV = \frac{K_1}{2} \cdot \left\{ \frac{2h^3}{3} \iint (\tilde{\nabla}^2 \xi)^2 dxdy \right.$$
$$\left. - 2h^2 \cdot \iint (\tilde{\nabla}^2 \xi) \cdot (\tilde{\nabla} \cdot \mathbf{f}) dxdy + 2h \cdot \iint (\tilde{\nabla} \cdot \mathbf{f})^2 dxdy \right\} + \frac{\alpha}{2} \iint \xi^2 dxdy \quad (21)$$

Curvature of the confining potential at its minimum $\alpha = \left. \frac{d^2W}{d\xi^2} \right|_{\xi=0}$ is calculated below using self-consistency procedure.

In Fourier space $\mathbf{q} = \{q_x, q_y\}$ the free energy functional (21) is written as:

$$F_v = \int_0^\infty \int_0^\infty (K_1 \cdot \frac{2h^3}{3} q^4 + \alpha) \cdot |\xi_q|^2 \frac{dq_x dq_y}{(2\pi)^2}$$
$$- \int_0^\infty \int_0^\infty K_1 \cdot h^2 \cdot i \cdot q^2 \cdot (\xi_q \cdot \mathbf{q} \cdot \mathbf{f}_q^* - \xi_q^* \cdot \mathbf{q} \cdot \mathbf{f}_q) \cdot \frac{dq_x dq_y}{(2\pi)^2} \quad (22)$$
$$+ \int_0^\infty \int_0^\infty 2 \cdot K_1 \cdot h \cdot |\mathbf{q} \cdot \mathbf{f}_q|^2 \frac{dq_x dq_y}{(2\pi)^2}$$

where $\mathbf{q}^2 = q_x^2 + q_y^2$.

We diagonalize the quadratic form in (22) with respect to $\xi_q$ and $\mathbf{q} \cdot \mathbf{f}_q$ by linear transformation:

$$\text{Re}\,\tilde{\xi}_q = \text{Re}\,\xi_q - \frac{3}{2h} \frac{q^2}{\tilde{q}^4} \text{Im}\,\mathbf{q} \cdot \mathbf{f}_q,$$

$$\text{Im}\,\tilde{\xi}_q = \text{Im}\,\xi_q + \frac{3}{2h} \frac{q^2}{\tilde{q}^4} \text{Re}\,\mathbf{q} \cdot \mathbf{f}_q, \quad (23)$$

where $\tilde{q}^4 = q^4 + \dfrac{\alpha}{K_b}$, $K_b = \dfrac{2h^3}{3}K_1$.

In the variables $\tilde{\xi}_q$ and $\mathbf{f}_q$ the energy functional (22) takes the form:

$$F_v = \int_0^\infty \int_0^\infty \frac{K_1}{2}\left\{\frac{4h^3}{3}\tilde{q}^4 \cdot |\xi_q|^2 + h\cdot\left(4 - 3\cdot\left(\frac{q}{\tilde{q}}\right)^4\right)\cdot|\mathbf{q}\cdot\mathbf{f}_q|^2\right\}\frac{dq_x dq_y}{(2\pi)^2}. \tag{24}$$

Using relations (23) and functional (24) we calculate the thermodynamical average $\langle|\xi_q|^2\rangle$:

$$\langle|\xi_q|^2\rangle = \frac{k_B T}{\dfrac{2h^3}{3}K_1 q^4 + \alpha} + \frac{3q^4 k_B T}{\left(\dfrac{2h^3}{3}K_1 q^4 + \alpha\right)\left(q^4 + \dfrac{6\alpha}{K_1 h^3}\right)} \tag{25}$$

here $k_B$ is Boltzmann's constant, T is temperature.

In the absence of interlayer slide only the first term in equation (25) would remain, as obtained in [11,14]. The second term in (25) signifies enhancement of the bending fluctuations caused by interlayer slide. The latter leads to relaxation of the lateral stresses (see Appendix A and Figure 3) and thus to a decrease of free energy of the bent membrane.

The mean-square fluctuations of bending amplitude are found as:

$$\langle\xi^2(\vec{r})\rangle = \int_0^\infty \langle|\xi_q|^2\rangle q\frac{dq}{2\pi} = \sqrt{\frac{3}{32}}\frac{k_B T}{\sqrt{\alpha K_1 h^3}}. \tag{26}$$

In the confined geometry the average bending amplitude is restricted to finite space *2d*, available between the walls (neglecting volume occupied by the membrane itself, i.e.: *h<< d*), thus providing the self-consistency condition for determination of the effective rigidity $\alpha$:

$$\langle\xi^2\rangle = \mu d^2, \tag{27}$$

where $\mu \leq 1$.

Substituting (26) into (27) we obtain a self-consistency solution for $\alpha$:

$$\alpha = \frac{(k_B T)^2}{16\cdot\mu^2 d^4 K_b}. \tag{28}$$

We also evaluate here the curvature of the confining potential, $\alpha_0$, for a unit bilayer (without interlayer slide). In this case the second term on the right hand side of (25) is zero and hence:

$$\alpha_0 = \frac{(k_B T)^2}{64\cdot\mu^2 d^4 K_b}. \tag{29}$$

Thus, interlayer slide results in considerable enhancement ($\alpha/\alpha_0=4$) of the curvature of confining potential.

## 4 Bilayer dynamics: viscous modes

To study the dynamical properties of the introduced model of bilayer membrane with interlayer slide we determine here equations of motion and find the eigenmodes of the membrane surrounded by water solution. We are interested in the behavior of membrane confined between parallel walls (see Figure 1).

Let a flat membrane lie in the (*x,y*)-plane with the normal pointed along z-axis. We treat each monolayer constituting the membrane as a (unit) two-dimensional condensed structure. We require the equilibrium between viscous stresses exerted on the membrane surface by water solution and membrane restoring force. We neglect inertial effects and introduce three sources of viscous dissipation: water and membrane viscosities and interlayer drag The force balance equations are expressed as:

$$-\frac{\delta F_s}{\delta \xi} + \Pi_{zz}(z=+0) - \Pi_{zz}(z=-0) = 0 \tag{30}$$

$$\frac{\delta F_s}{\delta f_x} - 2 \cdot \eta_m \cdot h \cdot \frac{\partial}{\partial t}\left(\tilde{\nabla}^2 f_x\right) + 2 \cdot b_s \cdot \frac{\partial f_x}{\partial t} - \Pi_{xz}(z=+0) - \Pi_{xz}(z=-0) = 0 \tag{31}$$

$$\Pi_{xz}(z=+0) - \Pi_{xz}(z=-0) = 0 \tag{32}$$

Here the fluid stress tensor is defined as: $\Pi_{ik} = -p \cdot \delta_{ik} + \eta_w \cdot \left(\frac{\partial v_i}{\partial x_k} + \frac{\partial v_k}{\partial x_i}\right)$, where *p* denotes hydrostatic pressure, $\bar{v}$ - velocity and $\eta_w$ - viscosity of water solution. The fluid stress tensor is evaluated at upper (*z=+0*) and lower (*z=-0*) membrane surfaces and carries the sign of the normal. The first term on the left hand side of (30) is the elastic restoring membrane force, which is balanced by normal to the membrane surface viscous stress of the fluid. Equation (31) represents force balance in lateral direction and contains the following contributions [3,8]: a) tangential traction on inter-layer surface due to monolayers differential flow; b) coherent surface flow of the monolayers as unit surfaces (with dynamic viscosity $\eta_m$,); c) viscous drag between monolayers (characterized by coefficient $b_s$) which arises at finite velocity of their mutual slide; d) traction of the surrounding fluid. Equation (32) accounts for the absence of total stretching forces exerted by water on the membrane since we discuss here only pure bending deformations of membrane, i.e. when total area dilation is zero.

Besides the balance equations (30)-(32), Navier Stockes equations for water solutions surrounding the membrane should be added. In the small velocity limit, treating fluid as incompressible and neglecting inertia, the "creeping flow" equations are written:

$$\nabla p = \eta_w \cdot \Delta v,$$

$$\nabla \cdot v = 0. \tag{33}$$

The non-slip boundary conditions at membrane-water interface provide the continuity of normal and tangential velocities of the fluid and the membrane:

$$\frac{\partial \xi}{\partial t} = v_z(\pm 0) \tag{34}$$

$$\frac{\partial f_i}{\partial t} = v_i(z = +0), \quad -\frac{\partial f_i}{\partial t} = v_i(z = -0), \; i = x, y. \tag{35}$$

Confinement between parallel walls at distance *2d* implies vanishing of water velocity (normal and tangential components) at the walls surfaces:

$$v_j(z = \pm d) = 0, \; j = x, y, z. \tag{36}$$

In order to find dispersion relation we make Fourier-transform of the free energy functional (21) and of the force balance and creeping flow equations. For this purpose the vibration is expanded in plane waves propagating along x direction. Then, free energy density, $F_s(q,\omega)$ takes the following form:

$$F_s(q,\omega) = \frac{K_1}{2} \cdot \left\{ \frac{4h^3}{3} \cdot q^4 \cdot |\xi_q|^2 - 2 \cdot h^2 \cdot (\xi_q \cdot f_q^* - \xi_q^* \cdot f_q) \cdot q^3 \cdot i + 4 \cdot h \cdot |f_q|^2 \cdot q^2 \right\} + \alpha \cdot |\xi_q|^2 \tag{37}$$

and

$$F_v = \frac{L_y}{(2\pi)^2} \int_{-\infty}^{+\infty} \int_0^{+\infty} F_s(q,\omega) dq d\omega, \tag{38}$$

where $L_y$ is system dimension along y-axis, and we have omitted index $\omega$ in the subscripts of the Fourier components.

Restoring membrane forces are given by functional derivatives of the free energy:

$$\frac{\delta F_s}{\delta \xi_q^*} = \left( K_1 \frac{2 \cdot h^3}{3} \cdot q^4 + \alpha \right) \cdot \xi_q + K_1 \cdot h^2 \cdot f_q \cdot q^3 \cdot i \tag{39}$$

$$\frac{\delta F_s}{\delta f_q^*} = -K_1 \cdot h^2 \cdot \xi_q \cdot q^3 \cdot i + 2 \cdot K_1 \cdot h \cdot q^2 \cdot f_q \tag{40}$$

Fourier transforms of creeping flow equations (33) for the components of water velocity and pressure $v_x = w_x(z) \cdot e^{iqx - i\omega t}$, $v_z = w_z(z) \cdot e^{iqx - i\omega t}$ and $p = p_q(z) \cdot e^{iqx - i\omega t}$ are written as:

$$iq \cdot w_x + \frac{\partial w_z}{\partial z} = 0,$$

$$iq \cdot p_q = \eta_w \cdot (-q^2 \cdot w_x + \frac{\partial^2 w_x}{\partial z^2}),$$

$$\frac{\partial p_q}{\partial z} = \eta_w \cdot (-q^2 \cdot w_z + \frac{\partial^2 w_z}{\partial z^2}). \tag{41}$$

We find the following solutions of differential equations (41) with normal velocity continuous at z=0, obeying also condition of zero lateral stretching force acting on the membrane (equation (32)) and condition $v_x(z = +0) = -v_x(z = -0)$ resulting from equation (35):

$$p_q = \begin{cases} z > 0: & 2\eta_w (C_1 \cdot e^{qz} + C_2 \cdot e^{-qz}) \\ z < 0: & 2\eta_w (-C_1 \cdot e^{-qz} - C_2 \cdot e^{qz}) \end{cases} \tag{42}$$

$$w_z = \begin{cases} z > 0: & [C_1 z + C_3] \cdot e^{qz} + [C_2 z + C_4] \cdot e^{-qz} \\ z < 0: & [-C_1 z + C_3] \cdot e^{-qz} + [-C_2 z + C_4] \cdot e^{qz} \end{cases} \quad (43)$$

$$w_x = \begin{cases} z > 0: & \left[ C_1 z + C_3 + \dfrac{C_1}{q} \right] \cdot ie^{qz} + \left[ -C_2 z - C_4 + \dfrac{C_2}{q} \right] \cdot ie^{-qz} \\ z < 0: & \left[ -C_2 z + C_4 - \dfrac{C_2}{q} \right] \cdot ie^{qz} + \left[ C_1 z - C_3 - \dfrac{C_1}{q} \right] \cdot ie^{-qz} \end{cases} \quad (44)$$

This solution maintains the symmetry relations compatible with the confined geometry:

$$w_x(z, x + \tfrac{\pi}{q}) = w_x(-z, x), \quad w_z(z, x + \tfrac{\pi}{q}) = -w_z(-z, x) \quad (45)$$

Physical meaning of (45), according to definitions given before (41), is that $x/z$ –component of water velocity around vibrating membrane behaves symmetrically/ antisymmetrically under simultaneous translation by half-period ($x \to x + \pi/q$) along the wave propagation direction $x$ and mirror reflection in the mid-plane between the confining walls ($z \to -z$).

Farther, we exclude unknown coefficients $C_2$, $C_4$ using stick boundary conditions at the walls (36). Then we substitute solutions in the form (42)-(44) into Fourier transformed force balance equations (30), (31) (exploiting (39), (40)) and into non-slip conditions (34), (35) at the water-membrane interface. Thus, finally we obtain algebraic system of four linear homogeneous equations for unknowns $C_1$, $C_3$, $\xi_q$ and $f_q$:

$$\xi_q \cdot \left[ -\frac{2 \cdot h^3}{3} \cdot K_1 \cdot q^4 - \alpha \right] + f_q \cdot \left[ -K_1 \cdot h^2 \cdot i \cdot q^3 \right]$$
$$+ C_1 \left[ -4 \cdot \eta_w \cdot 2 \cdot q^2 \cdot d^2 \cdot e^{2qd} \right] + C_3 \left[ 4 \cdot \eta_w \cdot q \cdot (1 + e^{2qd} - 2qd \cdot e^{2qd}) \right] = 0 \quad (46)$$

$$\xi_q \cdot \left( -h^2 \cdot K_1 \cdot q^3 \cdot i - 4 \cdot \eta_w \cdot q \cdot \omega \right) + f_q \cdot \left( 2 \cdot K_1 \cdot h \cdot q^2 - 2 \cdot \eta_m \cdot h \cdot q^2 \cdot i \cdot \omega - 2 \cdot b_s \cdot i \cdot \omega \right)$$
$$+ C_1 \left[ -4 \cdot \eta_w \cdot i \cdot (1 + e^{2qd} + 2qd \cdot e^{2qd}) \right] + C_3 \cdot \left[ -4 \cdot \eta_w \cdot i \cdot 2 \cdot q \cdot e^{2qd} \right] = 0 \quad (47)$$

$$i \cdot \omega \cdot \xi_q + 2qd^2 e^{2qd} \cdot C_1 + C_3 \cdot (1 + 2qd e^{2qd} - e^{2qd}) = 0 \quad (48)$$

$$\omega \cdot f_q + \frac{C_1}{q} \cdot (1 - e^{2qd} - 2qd e^{2qd} - 2q^2 d^2 e^{2qd}) + C_3 \cdot (1 - 2qd e^{2qd} - e^{2qd}) = 0 \quad (49)$$

Dispersion relation $\omega(q)$ is found by equating to zero the determinant of the system (46)-(49). The latter gives quadratic equation for $\omega(q)$, which results in two branches $\omega_1(q)$ and $\omega_2(q)$, see Figure 2. Two viscous modes: hydrodynamically damped bending mode and inter-monolayer slipping mode - mix and the power law $\omega(q)$ changes with wavelength of fluctuations. For pure bending deformation of the membrane there exist up to four hydrodynamic regimes (depending on the parameters of the system), separated by three crossover wave vectors.

We use the result (28) to estimate an upper limit, $q_0$, of the smallest $q$-interval where the eigenmodes are modified by the confining potential, i.e. where the induced rigidity term $\sim \alpha$ dominates over the bending term $\sim K_b q^4$ in (22) (and in the first bracket in equation (46)):

$$q \ll q_0 \equiv \left(\frac{\alpha}{K_b}\right)^{1/4} = \left(\frac{k_B T}{K_b}\right)^{1/2} \frac{1}{2d}\left(\frac{1}{\mu}\right)^{1/2}. \tag{50}$$

For typical value of bending rigidity at room temperature $K_b \sim 25 k_B T$ [2] $q_0 \sim \frac{0.1}{d}$. The second crossover wave vector, $1/d$, bounds the long wavelength regime where confinement of the surrounding water between the walls effects membrane dynamics. For $q \gg 1/d$ membrane behaves as in the bulk water solution. We assume that distance between confining walls is much greater than monolayer thickness ($2d/h \sim 10$). The crossover wave vector for the bulk fluid $q_1$ (see Appendix B) at given choice of parameters $h = 2 \cdot 10^{-7}$ cm, $\eta_w = 10^{-2}$ dyn·sec/cm$^2$, $b_s = 10^7$ dyn·sec/cm$^3$ acquires the value $q_1 = \frac{\eta_w}{b_s h^2} \sim 10^5$ cm$^{-1}$ and therefore obeys $q_1 \ll 1/d$. Thus, it does not influence dynamic behavior of the membrane in confined geometry. In the interval of still shorter wavelengths there is one more crossover wave vector: $q_2 = \sqrt{\frac{b_s}{\eta_m h}} \sim 10^7/\sqrt{2}$ cm$^{-1}$ ($\eta_m = 1$ dyn·sec/cm$^2$), which obeys $1/d \ll q_2$. Hence, we investigate four intervals of wave vector values: $q \ll q_0$, $q_0 \ll q \ll 1/d$, $1/d \ll q \ll q_2$ and $q_2 \ll q$.

For long wavelengths: $q \ll 1/d$, confinement between the walls modifies the bending mode with respect to membrane in the bulk solution (see Appendix B):

$$\omega_1^B = -i \cdot q^3 \cdot \frac{K_1 h^3}{24 \eta_w} \sim -i \cdot q^3 \cdot \frac{K_b}{\eta_w}, \tag{51}$$

and results either in $q^2$- or in $q^6$-dependences of $\omega_1$ instead of $q^3$-dependence of the "bulk" mode. For $q \ll q_0$ the bending mode becomes:

$$\omega_1 = -i \cdot q^2 \cdot \frac{\alpha d^3}{24 \eta_w}. \tag{52}$$

The mode $\omega_1(q)$ is driven by steric potential, characterized by curvature $\alpha$, and is damped by viscous losses in the surrounding fluid. For $q_0 \ll q \ll 1/d$ the hydrodynamically damped bending mode is given by:

$$\omega_1 = -i \cdot q^6 \cdot \frac{K_1 h^3 d^3}{144 \eta_w} \sim -i \cdot q^6 \cdot \frac{K_b d^3}{\eta_w}. \tag{53}$$

In this wave vector interval finite thickness, $d$, of water layers effectively enhances water viscosity from $\eta_w$ to $\eta_w/(dq)^3 \gg \eta_w$. Result (47) coincides (modulo numerical coefficient) with the damped vibration mode of erythrocyte walls consisting of two membranes, which comprise liquid between them [5].

Simultaneously, inter-monolayer slipping mode, $\omega_2(q)$, damped by viscous drag at the monolayer's mutual interface, remains unchanged by confinement (see Appendix B):

$$\omega_2 = -i \cdot q^2 \cdot \frac{K_1 h}{b_s} \sim -i \cdot q^2 \cdot \frac{K_A}{b_s} \tag{54}$$

For a membrane in the bulk solution the mixing of bending and slipping modes occurs at $q \approx q_1$, defined in Appendix B. The relative order of the parameters $q_1$, $1/d$, $q_2$ by increasing value depends on the choice of characteristic parameters of the system. Under our choice $q_1 \ll 1/d$ and the mixing of the modes is delayed up to $q \approx 1/d$, see Figure 2. We speculate that this happens because confinement hinders bending fluctuations and therefore bending mode remains slower than slipping mode up to $q \approx 1/d$.

In short-wavelength limit, $q \gg 1/d$, we recover, as expected, the result for a membrane in the bulk water. Confinement is not revealed in this case because membrane-induced vibrations of water decay exponentially before reaching the walls. Namely, for $q \gg 1/d$ the branch $\omega_2(q)$ corresponds now to bending mode damped by viscous losses in the surrounding fluid.

$$\omega_2^B = -i \cdot q^3 \cdot \frac{K_1 h^3}{6 \eta_w} \sim -i \cdot q^3 \cdot \frac{K_b}{\eta_w}. \tag{55}$$

The renormalized bending rigidity $\sim K_1 \cdot h^3$ arises for high frequency fluctuations (compare the numerical coefficients in (51) and in (55)) because bending mode is faster than interlayer slipping mode [8,9]; inter-layer slide leading to relaxation of lateral stresses in monolayers is retarded. In the interval $1/d \ll q \ll q_2$, the branch $\omega_1(q)$ becomes inter-layer slipping mode with renormalized area compression modulus (superscript [B] below indicates that solution coincides with the bulk water case):

$$\omega_1^B = -i \cdot q^2 \cdot \frac{K_1 h}{4 b_s} \sim -i \cdot q^2 \cdot \frac{K_A}{b_s}, \tag{56}$$

Finally, for $q \gg q_2$ the $\omega_1(q)$ mode is driven by (high frequency) effective rigidity $K_1$ and is damped by monolayer surface viscosity $\eta_m$, which dominates over interlayer drag as the monolayers are dynamically coupled:

$$\omega_1^B = -i \cdot \frac{K_1}{4 \eta_m}. \tag{57}$$

Viscous modes for a membrane in confined geometry obtained in this paper are in qualitative accord with the results for membrane bound to substrate [9]. We have not included the membrane tension into our free energy functional, because in the considered limit of small bending deformations of the bilayer, the term proportional to gradient of bending amplitude vanishes [14].

Dispersion relation for bilayer membrane in the bulk water based on our free energy functional (20) is derived in Appendix B and is also in accord with earlier results, obtained using density-difference model [8].

## 5 Conclusions

A novel free energy functional of bilayer fluid membrane derived in this paper reflects important physical properties of the membrane defining its dynamic behavior. The functional allows for two-dimensional liquid-crystalline structure of the membrane and weak adherence between the constituting it monolayers which results in their mutual slide under (bending) deformations. Our free energy functional contains three coupled fields parametrizing degrees of freedom related with bending of membrane, interlayer mutual slide and area dilation.

Using this functional we have calculated self-consistently the curvature of effective steric potential acting on the membrane between two parallel confining walls. We found that the curvature at the potential's minimum (located at the middle between the walls) is enhanced four times for a bilayer with interlayer slide in comparison with a unit membrane (with forbidden slide) of the same thickness. This increase can be ascribed to (partial) decrease of lateral stress in the bent membrane via interlayer slide. The relaxation of stresses effectively lowers the energetic "cost" of membrane bending and increases thermodynamic probability for conformations with greater bending amplitudes. This in turn amplifies steric repulsion.

We have also calculated the dispersion relations for a membrane confined between parallel walls. Our results are in qualitative accord with those for membrane bound to a substrate [9]. Confinement modifies viscous modes $\omega(q)$ at long wavelengths compared to the bulk water case. We have found four wave vector intervals separated by three characteristic wave vector values: $q_0 << 1/d << q_2$, defined in Section 4. The inverse of the half-distance $d$ between confining walls divides $q$-axis into two intervals with confined (q<<1/d) and bulk (q>>1/d) behavior, respectively. Wave vector $q_0$ delimits the interval of $q$-values, in which steric potential modifies spectrum of bending modes. In the interval $q_0 << q << 1/d$ we found $\omega(q) \propto q^6$ dependence of bending mode, similar to peristaltic modes of a soap film [5]. Unlike in [9], we do not obtain $\omega(q) \propto q^4$ dependence, because overall membrane tension is not included into our free energy functional. Since we consider the limit of small bending deformations of a flat bilayer the term proportional to gradient of bending amplitude vanishes [14]. In the interval $q >> 1/d$ confinement is not important since membrane-induced vibrations of water decay exponentially before reaching the walls. As in the bulk case, at $q > q_2$ the monolayer surface viscosity $\eta_m$ dominates over interlayer drag and the monolayers become dynamically strongly coupled.

Finally, we mention some limitations and possible improvements of our approach. Our functional respects reflectional symmetry of a flat bilayer and therefore implies that spontaneous curvature is zero. We assumed a thin-plate approximation for each monolayer with constant elastic moduli. In other words, we developed phenomenological effective media model. Hence, only fluctuations with wavelength larger than inter-molecular distance in lipid monolayer are considered. We have exploited smallness of bending-to-thickness ratio using linear approximation for the stress tensor. In the small bending approximation area dilation is decoupled from bending. In this paper we discussed only pure bending deformations, nevertheless, area dilation dynamics can also be studied using our functional. We found only damped eigenmodes of the membrane in confined water solution. The propagating modes will be considered elsewhere.

The authors are grateful to prof. Robijn Bruinsma for the formulation of the problem and to prof. Yu. A. Chizmadzhev and his coworkers for useful comments.

**Appendix A: Analytical solutions for axially symmetric case**

We can obtain analytical results describing the equilibrium shape of and mutual monolayer slide in the bilayer lipid membrane under constant external pressure for the cylindrically symmetric case. Consider a flat (unperturbed) circular membrane in the plane *(x,y)* of the radius *R*. We search for an equilibrium solution independent of the polar angle $\phi$:

$$\xi = \xi(r), \tag{A.1}$$

where *r* is the radial coordinate in the reference system with the origin situated at the center of the unperturbed membrane's mid-plane, and z-axis directed along the membrane's normal. Hence, the slide-functions take the form:

$$f_x(x,y) = f(r) \cdot \cos\phi, \quad f_y(x,y) = f(r) \cdot \sin\phi, \tag{A.2}$$

which then leads to the following expression for the radial component of the distortion field:

$$u_r(r,z) = -z \cdot \frac{\partial \xi(r)}{\partial r} + (\Theta(z) - \Theta(-z)) \cdot f(r). \tag{A.3}$$

Since the deformation is purely radial, the angular component of the distortion is zero: $u_\phi = 0$. The symmetry of the distortion fields (A.1), (A.2) permits to express the free energy density (11) in the cylindrical coordinates as follows:

$$F = \frac{K_1}{2} \cdot (u_{rr} + u_{\phi\phi})^2, \tag{A.4}$$

where

$$u_{rr} = \frac{\partial u_r}{\partial r}, \quad u_{\phi\phi} = \frac{u_r}{r} + \frac{1}{r} \cdot \frac{\partial u_\phi}{\partial \phi} = \frac{u_r}{r}. \tag{A.5}$$

Equilibrium state of the membrane under pressure is defined by the Euler-Lagrange equations, which are obtained by equating to zero the first variational derivatives of the elastic energy functional $F(\xi,f)$ with respect to the functions $\xi(r)$ and *f(r)* entering $u_{rr}$ and $u_{\phi\phi}$ in accord with (A.5) and (A.3):

$$\frac{\delta F_{sr}}{\delta \xi(r)} - 2 \cdot \pi \cdot r \cdot P = 0,$$

$$\frac{\delta F_{sr}}{\delta f(r)} = 0, \tag{A.6}$$

where $F_v = \int_0^R F_{sr} dr$, P is z-component of an external pressure difference applied to the opposite sides of the membrane.

Equations (A.6) can be decoupled by the introduction of the new unknown functions *p(r)* and *g(r)* instead of functions $\xi$ and *f*:

$$p = \frac{4 \cdot h}{3} \cdot \frac{\partial \xi}{\partial r} - 2 \cdot f, \quad g = h \cdot \frac{\partial \xi}{\partial r} - 2 \cdot f, \tag{A.7}$$

In the new basic set of functions $\{p,g\}$ equations (A.6) are transformed accordingly into the following form:

$$r^3 \cdot p''' + 2 \cdot r^2 \cdot p'' - r \cdot p' + p = P_1 \cdot r^3,$$
$$r^2 \cdot g'' + r \cdot g' - g = 0. \tag{A.8}$$

where $P_1 = \dfrac{2 \cdot P}{K_1 \cdot h^2}$.

Both equations in (A.8) belong to the Euler's class of equations and can be solved analytically using the transformation of the variable: $r = e^x$, where $-\infty < x < \infty$ is the new variable.

The following boundary conditions are imposed:

1) $(p''(r) \cdot r + p'(r) - p(r)/r)\big|_{r=0} = 0$ - the bending amplitude $\xi(r)$ is arbitrary at $r=0$;

2) $\xi(R) = 0$ - membrane is fixed at the edge (no vertical displacement);

3) $(p'(r) \cdot r + p(r))\big|_{r=R} = 0$ - zero torque at the membrane's edge;

4) $\partial\xi/\partial r\big|_{r=0} = 0$ the slope at the center is zero ;

5) $f(0) = 0$ - no inter-monolayer slide at the center (axial symmetry);

6) $g'(r) \cdot r + g(r)\big|_{r=R} = 0$ - the inter-monolayer slide at the edge is arbitrary. (A.9)

These conditions have transparent physical meaning. The conditions 1) and 3) in (A.9) originate from the expression for the variational derivative $\delta F_{sr}/\delta\xi$, and the condition 6) arises in the variational derivative $\delta F_{sr}/\delta f$; both derivatives include integration by parts in the segment $[0 \le r \le R]$. In particular, condition 1) is obtained by equating to zero the prefactor in front of $\delta\xi(r=0)$. Condition 3) is derived by equating to zero the prefactor in front of $\partial\xi/\partial r\big|_{r=R}$, which in turn corresponds to zero torque, $M$, at the membrane's edge (hence, membrane's slope at the edge is arbitrary):

$$M = K_1 \cdot \pi \cdot h^2 \cdot \left( \frac{4}{3} \cdot h \cdot \left( \frac{\partial^2 \xi}{\partial r^2} \cdot r + \frac{\partial \xi}{\partial r} \right) - 2 \cdot \left( \frac{\partial f}{\partial r} \cdot r + f \right) \right) \tag{A.10}$$

Condition 2) models the fixation of the membrane at the periphery. Condition 4) implies a smooth shape at the center of the curved membrane. The resulting solutions are:

$$\xi(r) = \frac{3 \cdot P}{32 \cdot K_1 \cdot h^3} \cdot \left( r^4 - 4 \cdot R^2 \cdot r^2 + 3 \cdot R^4 \right),$$

$$f(r) = \frac{3 \cdot P}{16 \cdot K_1 \cdot h^2} \cdot \left( r^3 - 2 \cdot R^2 \cdot r \right). \tag{A.11}$$

The bending amplitude $\xi(r) = u_z(r)$ is defined at the interface (mid-plane) of membrane and is z-independent (for the considered here small bending of membrane). The function f(r) characterizes the amplitude of mutual slide of the monolayers at the interface of membrane (z=0) (the total amplitude is given by $2 \cdot f$). As a result of this slide the bottom surface of the upper monolayer is compressed, and the top surface of the lower monolayer is expanded. In the present approximate approach $f$ is constant along the thickness (along $z$-axis) of the monolayers and depends on the position in the plane of the membrane. It is apparent from (A.11) that $f \sim h \cdot \xi/R$.

Substituting (A.11) in the expression (A.3) for the radial distortion $u_r$, we find:

$$u_r = \frac{3 \cdot P}{16 \cdot K_1 \cdot h^2} \cdot \left( -z \cdot \frac{2}{h} - \Theta(-z) + \Theta(z) \right) \cdot \left( r^3 - 2 \cdot R^2 \cdot r \right). \tag{A.12}$$

The radial stress component corresponding to the distortion given by (A.12) is readily found:

$$\sigma_{rr}(r,z) = \frac{3 \cdot P}{4 \cdot h^2} \cdot \left( -z \frac{2}{h} - \Theta(-z) + \Theta(z) \right) \cdot \left( r^2 - R^2 \right). \tag{A.13}$$

It is important to mention here that the lateral stress component $\sigma_{rr}$ in (A.13) proves to be independent of the elastic modulus ($K_1$) in our weak bending approximation. On the other hand, the distortion and slide fields and the strain tensor components depend on elastic modulus.

In the considered case of small bending amplitude there is no overall stretch of the deformed membrane (i.e. the pure bending takes place) and thus at any $r$:

$$\int_{-h}^{h} \sigma_{rr}(r) dz \sim \int_{-h}^{h} u_{rr}(r) + u_{\phi\phi}(r) dz = 0. \tag{A.14}$$

Fulfillment of this equality is guaranteed by the $z$-dependent factor in equation (A.13). The condition (A.14) is kept by the equality of the factors in front of $\Theta(+z)$ and $\Theta(-z)$ (i.e. the $f$-function is taken to be the same in both monolayers). Simultaneously, stretching deformation of the monolayers equals zero: $\mathbf{a} \equiv 0$ in the definitions of the distortion field components (see expression (14) in Section 3). In general, if one does not restrict the problem to the weak bending deformation and/or if there are additional forces acting in the lateral direction (stretching the membrane), one may introduce $\mathbf{a}(\mathbf{r}) \equiv 0$ or use two functions $f_1 \neq f_2$ in front of $\Theta(+z)$ and $\Theta(-z)$ respectively.

Results of the analytical solution of static equations in the cylindrically symmetric case are presented in Figure 3. Lateral stress $\sigma_{rr}(r,z)$ is shown for several values of $z$-coordinate for a bilayer with mutual interlayer slide (solid lines and dashed lines) and for a unit bilayer with forbidden slide, but of the same thickness $2h$ (dotted lines). Relaxation of lateral stresses in both monolayers is induced by mutual interlayer slide. The neutral (not stretched) surface at the interface of the membrane splits into two ones. Consequently, a neutral surface (with vanishing lateral stress) appears in the middle of each monolayer: at $z = +h/2$ (upper monolayer) and at $z = -h/2$ (lower monolayer), see dashed line. The monolayers are deformed as if they were disconnected, independent layers, but still adjusted to the same shape defined at their mutual interface inside the membrane. Therefore, the stress profiles along $z$-axis coincide with each other in both monolayers. As a result the stresses at the top and bottom external surfaces of the membrane ($z = \pm h$, solid lines) decrease two times with respect to the case without slide ($z = \pm h$, dotted lines). Simultaneously, as it follows from (A.13), the lateral stresses at the boundary $r=R$ turn to zero through the whole depth of the membrane: $\sigma_{rr}(R,z) = 0$, corresponding to the absence of the applied external stretching forces.

**Appendix B: Bilayer modes in the bulk water**

In order to test the relevance of our approach for description of dynamical properties of a bilayer, we rederive here the dispersion relation for a membrane in the bulk water solution using our free energy functional (20), introduced in Section 2. Our results are in accord with earlier ones, obtained for a membrane in the bulk fluid using curvature elastic model [5] and density-difference model [8].

For surrounding bulk fluid we search for the solution of creeping flow equations (33) (Section 4) satisfying the non-slip conditions at membrane-water interface (34)-(35). In addition, we impose the following boundary conditions for fluid velocity components $v_i$ :

$$v_j(z = \pm\infty) = 0, \quad j = x, y, z; \qquad (B.1)$$

which require the fluid velocity field vanish at large distances from the membrane.

As in Section 4, we expand vibrations in plane waves propagating along x-axis. We make Fourier transform of the free energy functional (20). The free energy density $F_s(q, \omega)$ is written as:

$$F_s(q,\omega) = \frac{K_1}{2} \cdot \left\{ \frac{4h^3}{3} \cdot q^4 \cdot |\xi_q|^2 - 2 \cdot h^2 \cdot (\xi_q \cdot f_q^* - \xi_q^* \cdot f_q) \cdot q^3 \cdot i + 4 \cdot h \cdot |f_q|^2 \cdot q^2 \right\} \qquad (B.2)$$

The components of water velocity and pressure in the form: $v_x = w_x(z) \cdot e^{iqx - i\omega t}$, $v_z = w_z(z) \cdot e^{iqx - i\omega t}$ and $p = p_q(z) \cdot e^{iqx - i\omega t}$ are substituted into Fourier transformed creeping flow equations (41) (see Section 4). The solutions of differential equations (41) satisfying boundary conditions (B.1), with normal velocity continuous at z=0, and obeying also condition of zero lateral stretching force acting on the membrane (equation (32) in Section 4) are the following:

$$p_q = \begin{cases} z > 0: & -2\eta_w \cdot C_1 \cdot e^{-qz} \\ z < 0: & 2\eta_w \cdot C_1 \cdot e^{qz} \end{cases} \qquad (B.3)$$

$$w_z = \begin{cases} z > 0: & [-C_1 z + C_3] \cdot e^{-qz} \\ z < 0: & [C_1 z + C_3] \cdot e^{qz} \end{cases} \qquad (B.4)$$

$$w_x = \begin{cases} z > 0: & \left[C_1 z - C_3 - \dfrac{C_1}{q}\right] \cdot i e^{-qz} \\ z < 0: & \left[C_1 z + C_3 + \dfrac{C_1}{q}\right] \cdot i e^{qz} \end{cases} \qquad (B.5)$$

Here constants $C_2$, $C_4$, which are present in the equations (42)-(44) in Section 4, turn to zero due to boundary conditions (B.1).

Unknown coefficients $C_1$, $C_3$ are determined from non-slip conditions (34), (35). Then, we substitute solutions (B.3)-(B.5) into Fourier transforms of force balance equations (30), (31) and obtain an algebraic system of two linear homogeneous equations for components $\xi_q$ and $f_q$:

$$\xi_q \cdot \left[ -\frac{2h^3}{3} \cdot K_1 \cdot q^4 + 4 \cdot \eta_w \cdot i \cdot q \cdot \omega \right] + f_q \cdot \left[ -K_1 \cdot h^2 \cdot i \cdot q^3 \right] = 0 \qquad (B.6)$$

$$\xi_q \cdot \left( -h^2 \cdot K_1 \cdot q^3 \cdot i \right) + f_q \cdot \left( 2 \cdot K_1 \cdot h \cdot q^2 - 2 \cdot i \cdot \omega \cdot (\eta_m \cdot h \cdot q^2 + 2 \cdot \eta_w \cdot q + b_s) \right) = 0 \qquad (B.7)$$

Equating to zero the determinant of the system (B.6)-(B.7) we obtain quadratic equation for $\omega(q)$, which results in two branches $\omega_1(q)$ and $\omega_2(q)$. There are three hydrodynamic regimes: $q<<q_1$, $q_1<<q<<q_2$ and $q_2<<q$, separated by crossover wave vectors $q_1$ and $q_2$ [8]:

$$q_1 = \frac{\eta_w}{b_s h^2}, \quad q_2 = \sqrt{\frac{b_s}{\eta_m h}}. \tag{B.8}$$

For long wavelengths, q<<q$_1$, the dispersion relations are given by:

$$\omega_1^B = -i \cdot q^3 \cdot \frac{K_1 h^3}{24\eta_w} \sim -i \cdot q^3 \cdot \frac{K_b}{\eta_w}. \tag{B.9}$$

$$\omega_2^B = -i \cdot q^2 \cdot \frac{K_1 h}{b_s} \sim -i \cdot q^2 \cdot \frac{K_A}{b_s}, \tag{B.10}$$

which describe respectively hydrodynamically damped bending mode, $\omega_1^B(q)$, and inter-monolayer slipping mode, $\omega_2^B(q)$, damped by viscous drag at the membrane mid-plane. Here superscript B is introduced to label membrane modes in the bulk fluid.

For wave vectors in the interval $q_1<<q<<q_2$ the bending and slipping modes mix [8]:

$$\omega_1^B = -i \cdot q^2 \cdot \frac{K_1 h}{4 b_s} \sim -i \cdot q^2 \cdot \frac{K_A}{b_s}, \tag{B.11}$$

$$\omega_2^B = -i \cdot q^3 \cdot \frac{K_1 h^3}{6\eta_w} \sim -i \cdot q^3 \cdot \frac{K_b}{\eta_w}. \tag{B.12}$$

The branch $\omega_2^B(q)$ corresponds now to bending mode damped by viscous losses in the surrounding fluid, and the branch $\omega_1^B(q)$ describes the damping of the slipping mode. The elastic moduli in (B.11) and (B.12) differ in general from that in (B.9) and (B.10), because high-frequency (bending) fluctuations occur at non-relaxed monolayer surface densities [8].

In short-wavelength limit, $q>>q_2$, we obtain:

$$\omega_1^B = -i \cdot \frac{K_1}{4\eta_m}, \tag{B.13}$$

$$\omega_2^B = -i \cdot q^3 \cdot \frac{K_1 h^3}{6\eta_w} \sim -i \cdot q^3 \cdot \frac{K_b}{\eta_w}. \tag{B.14}$$

The $\omega_1^B(q)$ mode is driven by (high frequency) effective rigidity $K_1$ and damped by monolayer surface viscosity $\eta_m$. Effective rigidity is induced by dynamic coupling of monolayers [3]. Monolayer surface viscosity overwhelms interlayer drag and becomes the main source of dissipation.

**Figure captions**

**Figure 1.** Membrane in the confined geometry. A bilayer membrane (each monolayer of thickness *h*) is placed in water solution between parallel walls separated by distance *2d*. The bending amplitude $\xi = u_z$ is defined at the mid-plane and is independent of the depth in the membrane. Inter-layer slide function **f** parameterizes position-dependent mutual slide of the monolayers at their interface.

**Figure 2.** Viscous modes of a bilayer membrane in water solution confined between parallel walls for the case of pure bending deformations. Damping rates $|\omega_i|$ (1/sec) are plotted as functions of dimensionless parameter (*qd*), where q is wave vector, *2d* is distance between the walls. Two branches 1 and 2 originate from bending and interlayer slide. The following values of parameters are used: $d=10^{-6}$ cm, $h=2\cdot10^{-7}$ cm, $\eta_w=10^{-2}$ dyn·sec/cm$^2$, $\eta_m=1$ dyn·sec/cm$^2$, $b_s=10^7$ dyn·sec/cm$^3$, $K_1=2\cdot10^8$ erg/cm$^3$.

**Figure 3.** The lateral stress $\sigma_{rr}$ normalized by $\dfrac{3\cdot P\cdot R^2}{4\cdot h^2}$ for various z-positions inside the membrane (z>0 for the upper monolayer, z<0 – for the lower) is plotted as the function of radial coordinate *r* (in dimensionless units). The solid lines show stresses in the upper and lower monolayers (the stress profiles along z-axis in both monolayers coincide due to interlayer slide, see text). The dashed line represents two neutral surfaces (at $z = \pm h/2$) in the lower und upper monolayers. The dotted lines (*z=h, z=-h*) characterize the stresses in the membrane of the same thickness *2h* with forbidden slide.

**Figure 1**

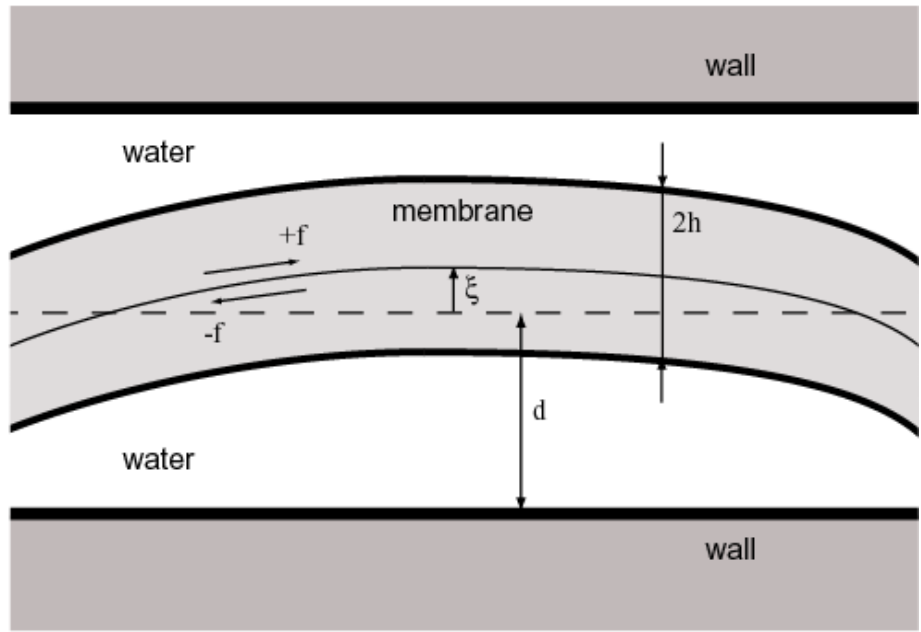

**Figure 2**

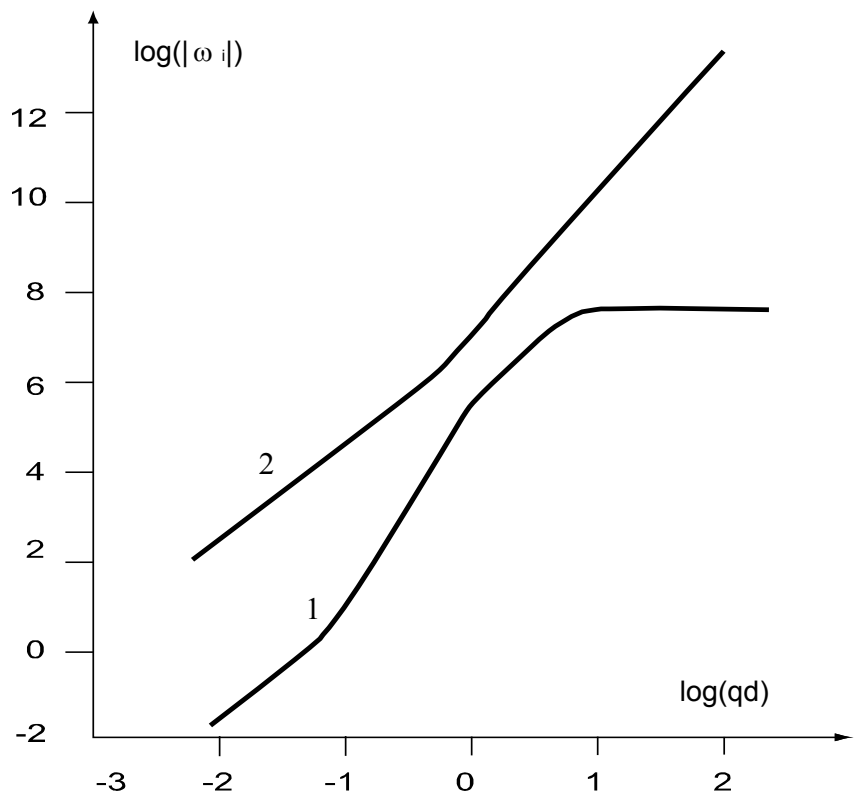

**Figure 3.**

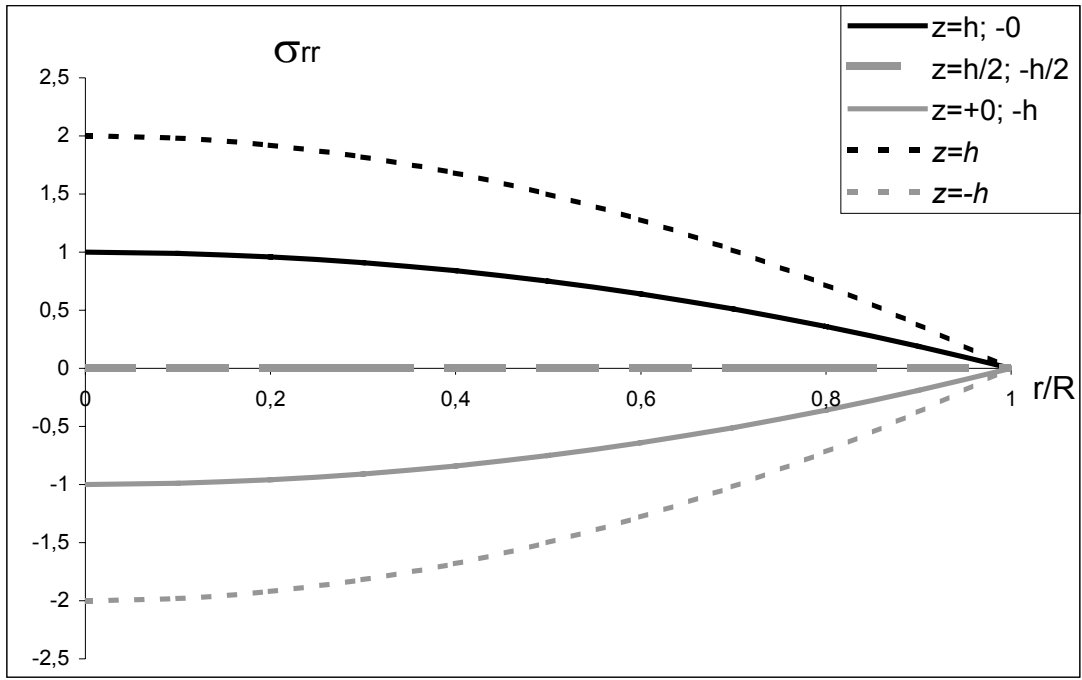